\documentclass[reprint,nofootinbib,preprintnumbers,amsmath,amssymb,onecolumn,unsortedaddress,superscriptaddress,twocolumn]{revtex4-1}
\usepackage[utf8]{inputenc}
\usepackage{graphicx, cancel, booktabs, tabularx, amsmath, amsfonts, amssymb, bm, hyperref, placeins}
\usepackage[normalem]{ulem}

\usepackage[table]{xcolor}

\renewcommand{\vec}[1]{\mathbf{#1}}

\newcolumntype{L}[1]{>{\raggedright\arraybackslash}p{#1}}

\newcolumntype{C}[1]{>{\centering\arraybackslash}p{#1}}

\newcolumntype{R}[1]{>{\raggedleft\arraybackslash}p{#1}}

\begin{document}
\title{Hierarchical symbolic regression for identifying key physical parameters correlated with bulk properties of perovskites}
\author{Lucas Foppa}
\affiliation{The NOMAD Laboratory at Fritz-Haber-Institut der Max-Planck-Gesellschaft, Faradayweg 4-6, D-14195 Berlin, Germany}
\affiliation{The NOMAD Laboratory at Humboldt-Universität zu Berlin, Zum Großen Windkanal 6, D-12489 Berlin, Germany}
\author{Thomas A. R. Purcell}
\affiliation{The NOMAD Laboratory at Fritz-Haber-Institut der Max-Planck-Gesellschaft, Faradayweg 4-6, D-14195 Berlin, Germany}
\affiliation{The NOMAD Laboratory at Humboldt-Universität zu Berlin, Zum Großen Windkanal 6, D-12489 Berlin, Germany}
\author{Sergey V. Levchenko}
\affiliation{Skolkovo Institute of Science and Technology, Moscow, Russia}
\author{Matthias Scheffler}
\affiliation{The NOMAD Laboratory at Fritz-Haber-Institut der Max-Planck-Gesellschaft, Faradayweg 4-6, D-14195 Berlin, Germany}
\affiliation{The NOMAD Laboratory at Humboldt-Universität zu Berlin, Zum Großen Windkanal 6, D-12489 Berlin, Germany}
\author{Luca M. Ghringhelli}
\affiliation{The NOMAD Laboratory at Fritz-Haber-Institut der Max-Planck-Gesellschaft, Faradayweg 4-6, D-14195 Berlin, Germany}
\affiliation{The NOMAD Laboratory at Humboldt-Universität zu Berlin, Zum Großen Windkanal 6, D-12489 Berlin, Germany}

\begin{abstract}
Symbolic regression identifies key physical parameters describing materials properties by uncovering correlations as nonlinear analytical expressions. However, the pool of expressions grows rapidly with complexity, compromising its efficiency. We tackle this challenge by a hierarchical approach: identified expressions are used as input parameters for obtaining more complex expressions. Crucially, this framework can transfer knowledge among properties, highlighting physical relationships. 
We demonstrate this strategy by using the \underline{s}ure-\underline{i}ndependence-\underline{s}creening-and-\underline{s}parsifying-\underline{o}perator (SISSO) approach to identify expressions correlated with the lattice constant and cohesive energy, which are then used to model the bulk modulus of $AB$O$_3$ perovskites.

\end{abstract}
\date{\today}
\maketitle
The identification of physical parameters that are correlated with materials properties or functions is a key step for understanding the underlying processes and accelerating the discovery of improved or even novel materials~\cite{Ghiringhelli-2015}. 
Because these parameters reflect the different processes that trigger, facilitate, or hinder a certain property or function, they might be referred to as \textit{materials genes}, in analogy to genes in biology.
Ideally, one would use physical models to describe the materials properties of interest~\cite{Reuter2005}. However, due to the intricate interplay of processes that might be responsible for a certain materials property, the explicit physical modelling might be unfeasible, or even inappropriate.
An alternative approach is to use artificial intelligence (AI) to uncover complex relationships by a data-centric concept.
Nevertheless, most widely used AI approaches require data sets that are much larger than those which are typically available in materials science, and only few AI methods incorporate the small-data aspect.\cite{Feng-2019,Batzner-2021,Debreuck-2021}
Furthermore, conventional AI produces black-box models that make it difficult to disentangle the contributions from the various input parameters and determine which underlying processes are the most important to optimize.
These problems are exacerbated for the typical scenario in which one is interested in finding  materials that exhibit an exceptional performance, for which only a few data points are available.

A possible avenue for linking physical reasoning and data-centric approaches is symbolic regression (SR).~\cite{Koza-1994,Wang-2019,Schmidt-2009} 
SR identifies nonlinear analytical expressions relating a target property to the key input parameters by using only small data sets. Thus, it is ideally suited for materials-science problems. These input parameters are physically meaningful quantities that are possibly related to the underlying processes governing the property. In the (initial) absence of understanding of the underlying processes, it is wiser to offer an extensive set of input parameters, also ones that are later revealed as irrelevant, rather than missing the important ones.
Traditionally, SR uses genetic-programming techniques to optimize the analytic expressions, which are combinations of the input parameters using mathematical operators such as addition, multiplication, exponentiation, etc., for a given problem~\cite{Koza-1994,Mueller-2014,Yuan-2017,Wang-2019,Udrescu-2019}.
These approaches randomly generate an initial population of possible expressions, and then stochastically apply genetic operators (e.g., mutation and crossover) until some optimal solution is found. 

More recently, the sure-independence-screening-and-sparsifying-operator (SISSO)~\cite{Ouyuang-2018,Ouyang-2019} approach was introduced for the identification of analytical expressions by applying the compressed sensing (CS) methodology ~\cite{Nelson-2013,Candes-2008} to SR.
The SISSO approach starts with the collection of physical input parameters, termed \textit{primary features}. Then, a more expansive pool of expressions is iteratively built by exhaustively applying mathematical operators to both the primary features and previously generated expressions. The number of recursive applications of the operators used to construct the pool of expressions is called the \textit{rung} ($q$) of the SR. Finally, CS is used to identify the $D$ expressions, which combined by weighting coefficients yield the best model for the property. In this step, an $\ell_0$-regularization is performed using only a subspace $S$ of the immense pool of generated expressions. This subspace is selected by the SIS procedure.
The outcome of the SISSO analysis is a low $D$-dimensional \textit{descriptor} vector containing, as components, the expressions selected from the pool of expressions.
A SISSO-derived model for a property $P$ has the form
\begin{equation} \label{eq:sisso_model}
    P^{\mathrm{SISSO}}=\sum_{i=0}^{D} c_i d_i,
\end{equation}
where $c_i$ are fitting coefficients, $d_i$ are the descriptor components, $D$ is the descriptor dimension. We also label the model components, $\alpha_i=c_i d_i$, which will be used for the construction of more complex models.
The primary features that appear in the (typically nonlinear) analytical expressions of the descriptor components reflect the materials genes of the property under investigation.

SR has been already used to model several materials properties and functions ~\cite{Bartel-2018, Xie-2019, Wang-2019, ouyang2019exploiting, cao2020, Foppa-2021, han2021, Ouyang-2021}. However, the combinatorial growth of the pool of possible expressions with respect to the number of primary features and to the number of times that the mathematical operators are applied can make an exhaustive search for the optimal descriptors impractical.
This is problematic because the relevant primary features are typically not known \textit{a priori} and one would like to initially offer as many as possible.
For addressing this challenge, we introduce a \textit{hierarchical} SR approach that enables an efficient identification of complex descriptors by keeping the number of expressions considered in the analysis at a manageable level. The foundation of this approach is the systematic re-feeding of expressions identified in one step as inputs for the identification of more complex expressions in subsequent steps. A crucial implication of this hierarchical framework is that it can be extended to transfer knowledge learned for one property to another one, thus also highlighting physical relationships between materials properties.

We demonstrate the hierarchical SR approach in the context of SISSO. Hierarchical SISSO (hiSISSO)  starts with an initial set of primary features, which is used to obtain an initial model for the property of interest. Then, the obtained model and its components ($P^{\mathrm{SISSO}}$ and $\alpha_i$, respectively), evaluated for all the materials in the data set, are added to the initial primary feature set. Finally, using this extended primary feature set, new, more complex models are obtained by applying SISSO for a second time. Models and components obtained for one (or more) property (properties) with SISSO can also be used to model a second, related property.

\begin{figure}
    \centering
    \includegraphics[width=0.50\textwidth]{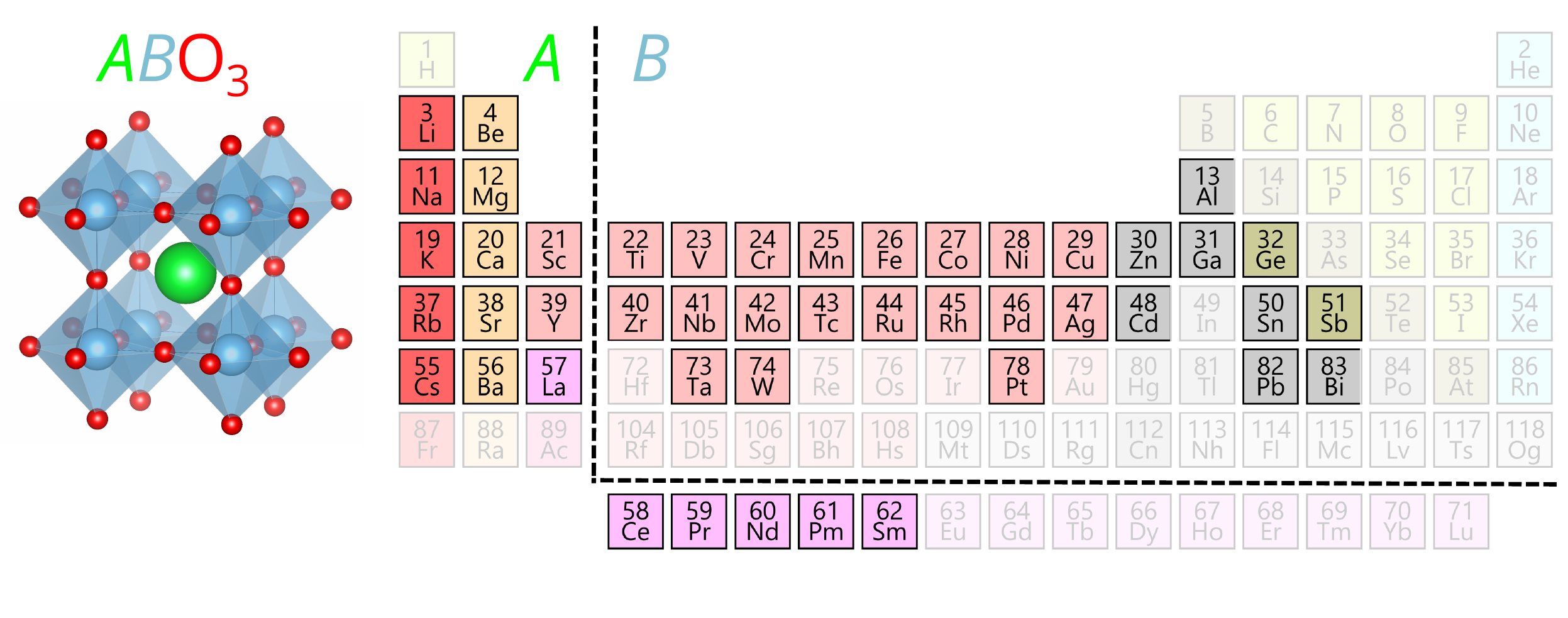}
    \caption{Materials space of $A$ and $B$ elements corresponding to the 504 cubic $AB$O$_3$ perovskites considered in the data set.}
    \label{fig:materials_space}
\end{figure}

In this work, we also introduce a new concept into SISSO, hereafter called ``multiple residuals'', which increases the algorithm's efficiency with respect to the size of subspaces needed for $\ell_0$-regularization, as shown in the ESI (Figure S2).
In the original SISSO algorithm~\cite{Ouyuang-2018}, the residual of the previously found model, $\vec{\Delta}_{D-1}^0$, i.e. difference between the vector storing the values of the property for each sample, $\vec{P}$,  and the estimates predicted by the by the ($D\! - \! 1$)-dimensional model ($\vec{\Delta}_{D-1}^0 = \vec{P}_{D-1} - \vec{P}$),  is used to calculate the projection score of the candidate features during the SIS step for the best $D$-dimensional model, $s_j^0 = R^2\left(\vec{\Delta}_{D-1}^0, \vec{d}_j\right)$. 
Here, $R$ is the Pearson correlation coefficient and $j$ corresponds to each element in the feature space. 
In the new implementation of SISSO\cite{Purcell-2021}, we extend the residual definition and use the best $r$ residuals to calculate the projection score: $\mathrm{max}\left(s_j^0, s_j^1, \ldots, s_j^{r-1} \right)$. 
This is used in the SIS procedure (see further details, including the choice of $r$, in the ESI). % 
The multiple-residual concept generalizes the descriptor identification step of SISSO, by using information from an ensemble of models to determine which features to add to the selected subspace.
By using the multiple-residual scheme with hiSISSO, we are able to expedite the search for the best models and considerably reduce (optimize) not only the overall size of the pool of expressions to be considered in the analysis, but also the size of the subspaces of expressions needed for the identification of the best descriptors. 

We demonstrate the capabilities of hiSISSO with two examples, i.e., by modeling the lattice constant ($a_0$) and bulk modulus of ($B_0$) of $AB$O$_3$ cubic perovskites. First, we identify models for the lattice constant ($a_0$) of each material. Then, we exploit the expressions identified for $a_0$ and cohesive energies ($E_0$) to improve the learning of the bulk moduli ($B_0$) of the perovskites. 
We consider 504 $AB$O$_3$ materials formed by the $A$ and $B$ elements indicated in Fig.~\ref{fig:materials_space}. 
The lattice constants, cohesive energies and bulk moduli were calculated using density-functional-theory (DFT) calculations with the PBEsol exchange-correlation functional.\cite{Csonka-2009} Further details and benchmarks of the calculation method are available in the ESI. The data set of calculated perovskite properties is available at the Novel-Materials Discovery (NOMAD) Repository \& Archive \cite{Data-set-NOMAD}.

Perovskites display a remarkable diversity of compositions and properties that make them interesting for very different functions and devices (e.g.,~\cite{Jena-2019,Hwang-2017}). We focus here on perovskite mechanical properties, specifically the equilibrium lattice constant, $a_0$, and the bulk modulus, $B_0$, the second derivative of the cohesive energy $E_0$ at $a_0$. Both quantities are correlated~\cite{Cohen-1988,Cohen-1985,Fischer-1993}, which has been described by Verma and Kummar (VK) for cubic perovskites:
\begin{equation}
B_0^{\mathrm{VK}}= C_0+C_1\frac{(n_A*n_B)^{C_2}}{(a_0)^{3.5}}.
    \label{eq:b0_ref}
\end{equation}
Here, $C_0$, $C_1$ and $C_2$ are fitted constants, and $n_A$ and $n_B$ are the expected oxidation state of the $A$ and $B$ species in the $AB$O$_3$ compound, as approximated by their group number on the periodic table. 
The approximation implies that all akali and alkaline earth metals will have an oxidation state of one and two, respectively, and all other $A$ elements will have an oxidation state of three. 
The oxidation state of the $B$ atom is then set to ensure all materials are charge-neutral, i.e., $n_B=6-n_A$.
The cohesive energy $E_0$ is defined as the energy per atom required to atomize the crystal.

As primary features, we use 23 properties related to the $A$ and $B$ elements of the $AB$O$_3$ perovskites, hereafter \textit{atomic features}. These atomic features only depend on the elements entering the composition of the materials, and not on their environment within the material: the radius of the valence-$s$ orbital and the radius of the highest-occupied orbital of the neutral atom ($r_s$ and $r_{\mathrm{val}}$, respectively), which represent the largest and smallest radii of the valence shell of an atom, respectively; the Kohn-Sham single-particle eigenvalue of the highest-occupied and lowest-unoccupied states ($\epsilon_\mathrm{H}$ and $\epsilon_\mathrm{L}$, respectively); the electron affinity ($EA$); the ionization potential ($IP$); and the electronegativity, ($EN$). Because $A$ and $B$ are present as cations in the perovskite structure, we also included the radius of the valence-$s$ orbital and the radius of the highest-occupied orbital of positively charged (+1 cations) atoms, denoted as $r_s^{\mathrm{cat}}$ and $r_{\mathrm{val}}^{\mathrm{cat}}$, respectively. All the above-mentioned atomic features were calculated using free (isolated) atoms with DFT-PBEsol\cite{Csonka-2009} (see further details in ESI) and are available in \cite{Feature-set-package}. The nuclear elemental charges $Z$ and the parameters $n_A$, $n_B$ and $n_A*n_B$ from Eq.~\ref{eq:b0_ref} were also included in our primary feature set. Note that $n_B$ depends on the formula of the perovskite, and, in particular, on the $A$ element. Let us also note that other choices of (elemental) primary features are possible (e.g., via pymatgen\cite{Ong-2013} or magpie\cite{Ward-2016}). 
We consider the following mathematical operators (where $\phi_i$ and $\phi_j$ are two arbitrary features): $\phi_i$, $\exp(\phi_i)$, $\exp(-\phi_i)$ $\ln(\phi_i)$, $\phi_i^{-1}$, $\phi_i^2$, $\phi_i^3$, $\phi_i^{1/2}$, $\phi_i^{1/3}$, $\phi_i+\phi_j$, $\phi_i-\phi_j$, $\phi_i*\phi_j$, $\frac{\phi_i}{\phi_j}$, $\left|\phi_i-\phi_j\right|$. This list of operators can be modified, reduced, or extended. 
In all cases, the units of the primary features are respected so that terms such as $\ln\left(r_\mathrm{s, A}\right)$ and $r_\mathrm{s, A} + Z_\mathrm{A}$ are not allowed.

\begin{figure*}
    \centering
    \includegraphics[width=0.9\textwidth]{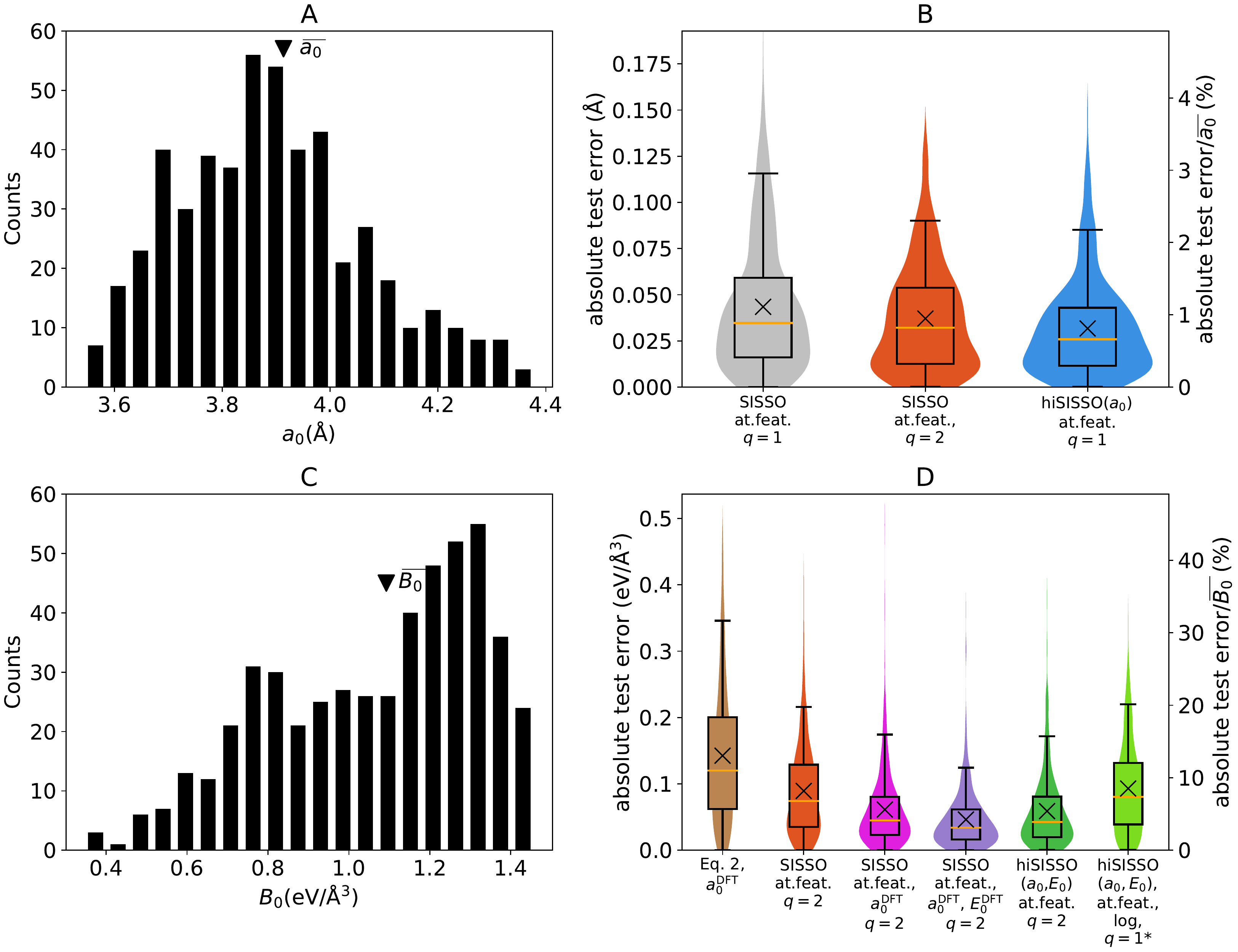}
    \caption{Identification of models for lattice constants ($a_0$) and bulk moduli ($B_0$) of cubic $AB$O$_3$ pervoskites using the hierarchical SISSO approach. (A) $a_0$ distribution over the entire data set of 504 materials. (B) Distribution of $a_0$ absolute test-set errors. (C) $B_0$ distribution over the entire data set. (D) Distribution of $B_0$ absolute test-set errors. The errors for the model of Eq.~\ref{eq:b0_ref} are also displayed in D. In B and D, the secondary axes (on the right) indicate the absolute test error as a fraction of the property average over the entire data set, the crosses indicate the mean absolute error, and the box plots show 0, 25, 50 (median, in orange), 75 and 95\%-iles of the test error distributions. In the figure labels, "at. feat." stands for atomic features and the star in $q=1$* indicates the reduced set of operators used in the log-regression approach.}
    \label{fig:results}
\end{figure*}

In order to evaluate the performance of our models, we randomly split the data set of 504 materials into five subsets. Four subsets are combined and used to train the models (training set) and the remaining subset is used to assess the performance (test set). The training set is used to determine the optimal model complexity with respect to its predictability via a 5-fold cross-validation (CV) scheme (see ESI). Within SISSO, the model complexity is controlled by the rung $q$ used to construct the pool of expressions and by the descriptor dimension $D$. Here we consider descriptors with $D=1$ up to $D=5$. Once the model complexity is determined by CV, a model is trained using all the materials of the training set at the optimal complexity. This model is used to predict the properties of the materials in the test set. Finally, the whole procedure is repeated five times, i.e., so that each of the five subsets is considered once as test set. We discuss the performance of the SISSO-derived models based on the distribution of absolute test errors across the 504 materials.

The distribution of $a_0$ values over the entire data set (504 materials) is shown in Fig.~\ref{fig:results}A. The absolute-test-error distributions associated to the models obtained with SISSO using rung $q=1$ and $q=2$ are shown as grey and red violin plots in Fig.~\ref{fig:results}B. The absolute-test-error distribution is shifted towards lower values when the rung increases from 1 to 2. This shows that the models become more accurate as the mathematical operators are applied for a second time in order to generate more complex expressions.
With our primary features and set of operators, rung 1 and 2 pools of features contain on the order of thousands and millions of elements, respectively. To demonstrate how complex descriptors can be found while keeping the number of considered expressions small, we collect the $a_0$, $q=1$ model and its components, and use them as new primary features, along with the atomic features, in a second step of SISSO application. In this second step, we also used $q=1$. We refer to the resulting models as hiSISSO($a_0$) in Fig.~\ref{fig:results}B, the parentheses indicating that the expressions describing $a_0$ identified in the first step are added to the primary feature set, along with the atomic features, in the second step. 
We note that some of the primary features might be correlated with each other. We do not exclude any of these features from our analysis because they could contain complementary information (e.g., the difference between two correlated features might not be correlated with any of the features individually). 

The absolute test errors associated to the hiSISSO($a_0$) models (Fig. ~\ref{fig:results}B, in blue) are lower compared to the errors of the $q=1$ models obtained with one-step application of SISSO (Fig. ~\ref{fig:results}B, in grey). Additionally, the performance of the hiSISSO($a_0$) models is superior compared with the SISSO approach with $q=2$ (Fig. ~\ref{fig:results}B, in red). 
These results show that hiSISSO provides a tractable way of increasing the effective rung - and thus the complexity - of a model at a tiny fraction of the computational cost required for a higher rung, since the pool of expressions that needs to be treated is three orders of magnitude smaller. Moreover, the hiSISSO approach generates more accurate $a_0$ models compared to the standard SISSO strategy using either rung 1 or 2.

In Fig. ~\ref{fig:results}, we note the presence of outliers for which the absolute test errors are high with respect to the distribution average. These data points are associated to materials with $A$ and/or $B$ elements which are significantly different compared to the $A$ and $B$ elements in the training sets. The detailed analysis of test errors is presented in ESI along with the discussion of a test set composed by materials containing chemical elements which were unseen during training.

We next address the bulk modulus of the perovskites. The distribution of values for this property over the entire data set is shown in Fig.~\ref{fig:results}C. The distribution of absolute test errors associated to Eq.~\ref{eq:b0_ref} (with $C_0$, $C_1$, and $C_2$ fitted to the training sets) is shown in brown in Fig.~\ref{fig:results}D as a baseline for evaluating the performance of the models derived by AI.  
For the SISSO analysis of bulk modulus, we consider rung $q=2$. 
The absolute-test-error distribution corresponding to the SISSO models obtained with the atomic features (Fig.~\ref{fig:results}D, in red) shows that this approach has an improved performance compared to the Eq.~\ref{eq:b0_ref}. 
By including the DFT-calculated lattice constant, $a_{0}^{\mathrm{DFT}}$, as well as $(a_{0}^{\mathrm{DFT}})^{-3.5}$ (as suggested by Eq.~\ref{eq:b0_ref}) as primary features along with the atomic features, the performance improves significantly (Fig.~\ref{fig:results}D, in magenta). If the DFT-calculated cohesive energy, $E_0^{\mathrm{DFT}}$, is further included as primary feature (Fig.~\ref{fig:results}D in violet), the errors get even smaller. This shows that $a_0$ and $E_{0}$ are both key parameters to describe the bulk modulus. 

The lattice constant and the cohesive energy provide necessary information to model the bulk modulus. However, the use of $a_{0}^{\mathrm{DFT}}$, $(a_{0}^{\mathrm{DFT}})^{-3.5}$, and $E_{0}^{\mathrm{DFT}}$ as primary features is inconvenient. In order to calculate $a_0$ and $E_{0}$ in DFT, one must perform a geometry relaxation, which is already majority of the work needed to calculate $B_0$ itself. To circumvent this issue, we offered, as primary features, the SISSO and hiSISSO models for $a_0$ and $E_{0}$ (see ESI) - as well as their components and the rescaled quantity $(a_0)^{-3.5}$ - instead of the DFT-calculated quantities. In this analysis, the atomic features are kept in the primary feature set. We indicate the resulting $B_0$ models by hiSISSO($a_0,E_{0}$) in Fig.~\ref{fig:results}D (dark green). 
By using the hiSISSO($a_0,E_{0}$) approach, the test errors are significantly reduced compared to the one-step application of SISSO to the atomic features. Indeed, the model performance gets closer to that of the models obtained using the DFT-calculated parameters $a_{0}^{\mathrm{DFT}}$ and $E_0^{\mathrm{DFT}}$, even though the hiSISSO($a_0,E_{0}$) models depend only on the atomic features, which makes it useful to search for new materials. These results demonstrate the potential of hiSISSO to transfer information among materials properties, thus circumventing the use resource-consuming primary features. 

We then exploit a $B_0$ model obtained by the hiSISSO($a_0,E_{0}$) approach, trained using the entire data set of 504 $AB$O$_3$ materials, for the screening of new materials (see details in ESI). We evaluate $7\, 308$ single ($AB$O$_3$) and double perovskite compositions of the type $A_2BB'$O$_6$ constructed from all the $A$ and $B$ elements in the initial data set (Fig.~\ref{fig:materials_space}). Then, we look at the materials with the lowest predicted $B_0$ values, since they are scarce in the training set. This situation corresponds to the typical scenario in materials discovery, in which the behavior of interest is associated to only few of the available observations. 
Among the 10 materials with the lowest $B_0$ predicted by the hiSISSO approach, we identify the double perovskites Cs$_2$ZnBiO$_6$, Cs$_2$CdBiO$_6$, Cs$_2$CdPbO$6$, Cs$_2$ZnPbO$_6$, Cs$_2$ZnCdO$_6$, Rb$_2$ZnBiO$_6$, and Rb$_2$CdBi$_6$, with predicted $B_0$ in the range 0.49-0.53  eV/\AA$^3$. The properties of these materials were evaluated explicitly by further DFT calculations and they were confirmed as highly compressible perovskites, with DFT-calculated $B_0$ of 0.45, 0.45, 0.46, 0.45, 0.46, 0.60, and 0.41 eV/\AA$^3$, respectively. 
The root-mean-squared error calculated on the 10 materials with the lowest predicted $B_0$ is 0.081 eV/\AA$^3$. 
By recalling that the model was trained on simpler single perovskites, its predictive ability beyond the training region is remarkable. 
Moreover, only 8 materials, out of the 504 used for training, present $B_0<0.50$ eV/\AA$^3$. The minimum $B_0$ value in the training set is 0.37 eV/\AA$^3$ for the CsCdO$_3$ material. The hiSISSO-predicted $B_0$ for this materials is 0.51 eV/\AA$^3$.

Finally, we applied a logarithm transformation to identify, with hiSISSO, a power-law-type expression for $B_0$. For the purpose of obtaining an equation in the spirit of Eq.~\ref{eq:b0_ref}, we offered the atomic features and the SISSO $q=2$ models for $a_0$ and $E_{0}$, denoted $a_0^{\mathrm{SISSO}(q=2)}$ and $E_0^{\mathrm{SISSO}(q=2)}$, respectively, as primary features. The components of these models are also included in the primary feature set. Here, we use $q=1$ with a reduced mathematical operator set containing only the operators addition and subtraction.
The best model identified using the entire data set of 504 materials at the optimal dimensionality identified via CV ($D=3$, see ESI) is
\begin{equation} \label{eq:b0_log_reg}
B_0^{\mathrm{hiSISSO}}=2.99 \frac{(IP_B-EA_B)^{0.419}(E_0^{\mathrm{SISSO}(q=2)})^{0.964}}{(a_0^{\mathrm{SISSO}(q=2)}-5.09*10^{-4}\frac{EA_B n_A}{|r_{s,B}^{\mathrm{cat}}-r_{s,B}|})^{2.75}}.
\end{equation}
The equations for $a_0^{\mathrm{SISSO}(q=2)}$ and $E_0^{\mathrm{SISSO}(q=2)}$ are shown in ESI.
SISSO selects $IP_B$, $EA_B$, $r_{s,B}^{\mathrm{cat}}$, $r_{s,B}$, $a_0^{\mathrm{SISSO}}$, and $E_0^{\mathrm{SISSO}}$ as the key parameters correlated with $B_0$. Therefore, SISSO recovers the parameters $a_0$ and $n_A$, which also enter Eq.~\ref{eq:b0_ref}.
However, the description of $B_0$ provided by Eq.~\ref{eq:b0_log_reg} goes beyond the empirical model, since the log-regression models provides a significantly better performance (light green in Fig.~\ref{fig:results}D) compared to Eq.~\ref{eq:b0_ref}. This analysis illustrates the potential of hiSISSO to revisit and improve models derived from physical arguments based on a data-centric approach.

In this work, we introduced a hierarchical SR framework to efficiently address complex materials properties and functions. This approach provides the key physical parameters reflecting the underlying processes responsible for the behavior of interest, while increasing the performance of SR models.  
The analysis described in this publication can be found in a Jupyter notebook at the NOMAD AI Toolkit\cite{Notebook-NOMAD}, where it can be reproduced and modified directly in a web browser. 

This work was funded by the NOMAD Center of Excellence (European Union’s Horizon 2020 research and innovation program, grant agreement Nº 951786), the ERC Advanced Grant TEC1p (European Research Council, grant agreement Nº 740233), and the project FAIRmat (FAIR Data Infrastructure for Condensed-Matter Physics and the Chemical Physics of Solids, German Research Foundation, project Nº 460197019). T.P. would like to thank the Alexander von Humboldt Foundation for their support through the Alexander von Humboldt Postdoctoral Fellowship Program. L.F. acknowledges the funding from the Swiss National Science Foundation, postdoc mobility grant P2EZP2\_181617.

\bibliography{main}

\begin{thebibliography}{10}

\bibitem{Ghiringhelli-2015}
Luca~M. Ghiringhelli, Jan Vybiral, Sergey~V. Levchenko, Claudia Draxl, and
  Matthias Scheffler.
\newblock Big data of materials science: Critical role of the descriptor.
\newblock {\em Physical Review Letters}, 114(10):105503, 2015.

\bibitem{Reuter2005}
Karsten Reuter, Catherine Stampf, and Matthias Scheffler.
\newblock \textit{Ab Initio} atomistic thermodynamics and statistical mechanics
  of surface properties and functions.
\newblock In Sidney Yip, editor, {\em Handbook of Materials Modeling: Methods},
  pages 149--194. Springer Netherlands, Dordrecht, 2005.

\bibitem{Feng-2019}
Shuo Feng, Huiyu Zhou, and Hongbiao Dong.
\newblock Using deep neural network with small dataset to predict material
  defects.
\newblock {\em Materials \& Design}, 162:300--310, 2019.

\bibitem{Batzner-2021}
Simon Batzner, Albert Musaelian, Lixin Sun, Mario Geiger, Jonathan~P. Mailoa,
  Mordechai Kornbluth, Nicola Molinari, Tess~E. Smidt, and Boris Kozinsky.
\newblock E(3)-equivariant graph neural networks for data-efficient and
  accurate interatomic potentials.
\newblock {\em arXiv preprint:2101.03164}, 2021.

\bibitem{Debreuck-2021}
Pierre-Paul~De Breuck, Geoffroy Hautier, and Gian-Marco Rigagnese.
\newblock Materials property prediction for limited datasets enabled by feature
  selection and joint learning with modnet.
\newblock {\em npj computational materials}, 7:83, 2021.

\bibitem{Koza-1994}
John~R. Koza.
\newblock Genetic programming as a means for programming computers by natural
  selection.
\newblock {\em Statistics and Computing}, 4:87--112, 1994.

\bibitem{Wang-2019}
Yiqun Wang, Nicholas Wagner, and James~M. Rondinelli.
\newblock Symbolic regression in materials science.
\newblock {\em MRS Communications}, 9(3):793--805, 2019.

\bibitem{Schmidt-2009}
Michael Schmidt and Hod Lipson.
\newblock Distilling free-form natural laws from experimental data.
\newblock {\em Science}, 324(5923):81, 2009.

\bibitem{Mueller-2014}
Tim Mueller, Eric Johlin, and Jeffrey~C. Grossman.
\newblock Origins of hole traps in hydrogenated nanocrystalline and amorphous
  silicon revealed through machine learning.
\newblock {\em Physical Review B}, 89(11):115202, 2014.

\bibitem{Yuan-2017}
Fenglin Yuan and Tim Mueller.
\newblock Identifying models of dielectric breakdown strength from
  high-throughput data via genetic programming.
\newblock {\em Scientific Reports}, 7(1):17594, 2017.

\bibitem{Udrescu-2019}
Silviu-Marian Udrescu and Max Tegmark.
\newblock {AI} {F}eynman: a physics-inspired method for symbolic regression.
\newblock {\em arXiv preprint:1905.11481}, 2019.

\bibitem{Ouyuang-2018}
Runhai Ouyang, Stefano Curtarolo, Emre Ahmetcik, Matthias Scheffler, and
  Luca~M. Ghiringhelli.
\newblock \textsc{SISSO}: A compressed-sensing method for identifying the best
  low-dimensional descriptor in an immensity of offered candidates.
\newblock {\em Physical Review Materials}, 2(8):083802, 2018.

\bibitem{Ouyang-2019}
Runhai Ouyang, Emre Ahmetcik, Christian Carbogno, Matthias Scheffler, and
  Luca~M. Ghiringhelli.
\newblock Simultaneous learning of several materials properties from incomplete
  databases with multi-task \textsc{SISSO}.
\newblock {\em Journal of Physics: Materials}, 2(2):024002, 2019.

\bibitem{Nelson-2013}
Lance~J. Nelson, Gus L.~W. Hart, Fei Zhou, and Vidvuds Ozoliņš.
\newblock Compressive sensing as a paradigm for building physics models.
\newblock {\em Physical Review B}, 87(3):035125, 2013.

\bibitem{Candes-2008}
E.~J. Candes and M.~B. Wakin.
\newblock An introduction to compressive sampling.
\newblock {\em IEEE Signal Processing Magazine}, 25(2):21--30, 2008.

\bibitem{Bartel-2018}
Christopher~J. Bartel, Samantha~L. Millican, Ann~M. Deml, John~R. Rumptz,
  William Tumas, Alan~W. Weimer, Stephan Lany, Vladan Stevanović, Charles~B.
  Musgrave, and Aaron~M. Holder.
\newblock Physical descriptor for the gibbs energy of inorganic crystalline
  solids and temperature-dependent materials chemistry.
\newblock {\em Nature Communications}, 9(1):4168, 2018.

\bibitem{Xie-2019}
S.~R. Xie, G.~R. Stewart, J.~J. Hamlin, P.~J. Hirschfeld, and R.~G. Hennig.
\newblock Functional form of the superconducting critical temperature from
  machine learning.
\newblock {\em Phys. Rev. B}, 100:174513, Nov 2019.

\bibitem{ouyang2019exploiting}
Runhai Ouyang.
\newblock Exploiting ionic radii for rational design of halide perovskites.
\newblock {\em Chemistry of Materials}, 32(1):595--604, 2019.

\bibitem{cao2020}
Guohua Cao, Runhai Ouyang, Luca~M Ghiringhelli, Matthias Scheffler, Huijun Liu,
  Christian Carbogno, and Zhenyu Zhang.
\newblock Artificial intelligence for high-throughput discovery of topological
  insulators: The example of alloyed tetradymites.
\newblock {\em Physical Review Materials}, 4(3):034204, 2020.

\bibitem{Foppa-2021}
Lucas Foppa, Luca~M. Ghiringhelli, Frank Girgsdies, Maike Hashagen, Pierre
  Kube, Michael Hävecker, Spencer~J. Carey, Andrey Tarasov, Peter Kraus, Frank
  Rosowski, Robert Schlögl, Annette Trunschke, and Matthias Scheffler.
\newblock Materials genes of heterogeneous catalysis from clean experiments and
  artificial intelligence.
\newblock {\em MRS Bull.}, 46:1016–1026, Nov 2021.

\bibitem{han2021}
Zhong-Kang Han, Debalaya Sarker, Runhai Ouyang, Aliaksei Mazheika, Yi~Gao, and
  Sergey~V Levchenko.
\newblock Single-atom alloy catalysts designed by first-principles calculations
  and artificial intelligence.
\newblock {\em Nature communications}, 12(1):1--9, 2021.

\bibitem{Ouyang-2021}
Ning He, Runhai Ouyang, and Quan Qian.
\newblock Learning interpretable descriptors for the fatigue strength of
  steels.
\newblock {\em AIP Advances}, 11(3):035018, 2021.

\bibitem{Purcell-2021}
Thomas A.~R. Purcell, Matthias Scheffler, Christian Carbogno, and Luca~M.
  Ghiringhelli.
\newblock \textsc{SISSO}++: A \textsc{C}++ implementation of the sure
  independence screening and sparsifying operator approach.
\newblock {\em J. Open Source Softw.}, 2021.

\bibitem{Csonka-2009}
Gábor~I. Csonka, John~P. Perdew, Adrienn Ruzsinszky, Pier H.~T. Philipsen,
  Sébastien Lebègue, Joachim Paier, Oleg~A. Vydrov, and János~G. Ángyán.
\newblock Assessing the performance of recent density functionals for bulk
  solids.
\newblock {\em Physical Review B}, 79(15):155107, 2009.

\bibitem{Data-set-NOMAD}
\textsc{DOI}:10.17172/\textsc{NOMAD}/2022.02.21-3.

\bibitem{Jena-2019}
Ajay~Kumar Jena, Ashish Kulkarni, and Tsutomu Miyasaka.
\newblock Halide perovskite photovoltaics: Background, status, and future
  prospects.
\newblock {\em Chemical Reviews}, 119(5):3036--3103, 2019.

\bibitem{Hwang-2017}
Jonathan Hwang, Reshma~R. Rao, Livia Giordano, Yu~Katayama, Yang Yu, and Yang
  Shao-Horn.
\newblock Perovskites in catalysis and electrocatalysis.
\newblock {\em Science}, 358(6364):751, 2017.

\bibitem{Cohen-1988}
Marvin~L. Cohen.
\newblock Theory of bulk moduli of hard solids.
\newblock {\em Materials Science and Engineering: A}, 105-106:11--18, 1988.

\bibitem{Cohen-1985}
Marvin~L. Cohen.
\newblock Calculation of bulk moduli of diamond and zinc-blende solids.
\newblock {\em Phys. Rev. B}, 32:7988--7991, Dec 1985.

\bibitem{Fischer-1993}
George~J. Fischer, Zichao Wang, and Shun-ichiro Karato.
\newblock Elasticity of \textsc{C}a\textsc{T}i\textsc{O}$_3$,
  \textsc{S}r\textsc{T}i\textsc{O}$_3$ and \textsc{B}a\textsc{T}i\textsc{O}$_3$
  perovskites up to 3.0 \textsc{G}pa: the effect of crystallographic structure.
\newblock {\em Physics and Chemistry of Minerals}, 20(2):97--103, 1993.

\bibitem{Feature-set-package}
Aakash Naik and Thomas A.~R. Purcell.
\newblock https://gitlab.mpcdf.mpg.de/nomad-lab/atomic-features-package.

\bibitem{Ong-2013}
Shyue~Ping Ong, William~Davidson Richards, Anubhav Jain, Geoffroy Hautier,
  Michael Kocher, Shreyas Cholia, Dan Gunter, Vincent~L. Chevrier, Kristin~A.
  Persson, and Gerbrand Ceder.
\newblock Python materials genomics (pymatgen): A robust, open-source python
  library for materials analysis.
\newblock {\em Computational Materials Science}, 68:314 -- 319, 2013.

\bibitem{Ward-2016}
Logan Ward, Ankit Agrawal, Alok Choudhary, and Christopher Wolverton.
\newblock A general-purpose machine learning framework for predicting
  properties of inorganic materials.
\newblock {\em npj Computational Materials}, 2:2057--3960, 2016.

\bibitem{Notebook-NOMAD}
https://nomad-lab.eu/aitoolkit/hierarchical\_sisso.

\end{thebibliography}
\bibliographystyle{unsrt}

\end{document}